# Quantified-Self 2.0: Using Context-Aware Services for Promoting Gradual Behaviour Change

Li Guo
School of Physical Science and Computing,
The University of Central Lancashire
lguo@uclan.ac.uk

**ABSTRACT**
The recent development of smartphone and wearable sensor technologies enable general public to carry self-tracking tasks more easily. Much work have been devoted to life data collection and visualisation to help people with better self-understanding. We believe that although (self-awareness/knowledge discovery is an important aspect of personal informatics, knowledge maintenance is more, or at least equally, important. In this paper, we propose a proactive approach that uses the knowledge mined from people's activity data to nudge them towards a good lifestyle (better knowledge maintenance). For demonstration purpose, a trial study was designed and implemented for good sleep maintenance. In the study, we first use smartphones as activity trackers to collect various features in a non-intrusive manner. We then use those data to learn users' activity patterns, including daily step amount, app usages, bedding time, wakeup time and sleep duration. Subsequently, we analyse correlations that may have positive or negative impact on users' sleep qualities and finally we designed and implemented three proactive services that are able to generate customised advices in the "right" context to nudge users towards a better life style. The experiments results are positive showing that with the use of the proposed services 1> daily step amount have been increased by 3.03% on average in a 10 days study and 2> sleep durations are increased by 7% for two subjects.

**INTRODUCTION**
Along with the fast development of smartphone and wearable sensor technologies, people now are able to collect and gain access to many sources of data about their daily activities and lifestyle patterns. As a consequence, an interesting concept-quantified-self (QS) has been proposed and intensively developed in the last few years. QS is defined as any individual engaged in the self-tracking of any kind of personal activity/lifestyle information and/or their associated context information. New knowledge about individuals is revealed from the collected data and people are then able to reflect upon it. Nowadays, a wide range of tools or devices for such purposes are available in the market such as Fitbit pedometers, Nike+, Jawbone UP fitness trackers, Apple watch, and etc. Also, almost all the solution providers pair their devices or mobile apps with web portals for data aggregation, statistical visualisation. In addition, there are some proposals and work aiming on fusing data from different aspects (activities, exercises, work calendar, sleep pattern, food intake, vital signs and etc.) of our life and performing correlations amongst those data. Such work could be very helpful on revealing some hidden knowledge that people may not be aware of (Knowledge Discovery) [1].

Although some of the existing work has shown their interestingness to be good starting points, making QS to be adopted by a wider public group still faces few main challenges [2] [3] [4]. First of all, the current QS adopters are mainly people who already have strong desire to improve their lifestyles or try to reduce some of the healthy issues that they are experiencing such as poor sleep [5]. Majorities of general public are still not fully convinced or simply cannot be bothered with QS due to low awareness of how it would help with their long term healthy status [6]. Secondly, although the QS technologies provide more intuitive solutions for knowledge discovery, how such knowledge can be understood properly and how people could take further actions in order to benefit from those remain as unsolved issues [3]. In other words, even if people are clearly aware of all potential problems they are having or will have; there is still a big question mark on whether they are "ABLE" to do what they are "SUPPOSED" to. There are many life constraints which stop them from engaging with activities that are helpful for good lifestyles. Such constraints include time, workload, living expenses, social responsibilities, unhealthy habits developed already and many more [3]. Such life constraints can be termed as paternalism [7]. Last but not least, many of the current QS solutions work on a statistical basis. QS pioneers, more or less, have to have the technical skills that are required to carry the QS tasks (using wearable devices; self-interpreting statistical results aggregated from their data; looking at the charts and understand their meanings and etc.). A lot of these tasks are time consuming and require constant efforts. As generally QS only benefits people in a long run, it's difficult to maintain a high initiative to keep people doing these tasks if everything is based on a "good will" and cannot show its effectiveness in a relatively short term [8]. Similar phenomena has been researched and discussed in Economics as "*hyperbolic discounting*" [9].

To address the above challenges, methods that help people initiatively engage with QS technologies and constantly take necessary actions in a long run (Knowledge Maintenance) should be studied and deployed. In this paper, we present a set of smart context-aware services that uses the knowledge mined from people's activity data to help them maintain a good lifestyle. For demonstration purpose, a trial study focusing on good sleep maintenance is designed. In the study, we first use smartphones as activity trackers to collect various features in a non-intrusive manner. We then use those data to learn users' activity patterns, including daily step amount, app usages, bedding time, wakeup time and sleep duration. Subsequently, we analyse correlations that may have positive or negative impact on users' sleep qualities and finally we designed and implemented three proactive services that are able to generate customised advices in the "right" context to nudge users towards a better life style.

## RELATED WORK

**Self-Tracking Devices and Applications**
More recently, along with the development of smartphones and wearable devices, the concept of QS has been widely embodied in the design of sensing and monitoring applications because of its effectiveness that leads to increased self-health awareness and behaviour change. As sensors have become smaller and can be better integrated with smartphones, it is more convenient for people to track numerous types of lifestyle data. Realising the power of self-tracking in activating health behaviour change, automated sensing or smartphone based manual tracking features are often used in designing self-monitoring technology. Within the personal informatics domain, researchers and companies designed technology for tracking physical fitness [10] [11] [12] [13], sleep [14] [15], ECG/heart rate [16] [17], diabetes [18], blood pressure [19] and many more.

Most of the solutions work in a reactive/ obtrusive manner. For example, fitness devices can track sleep/sleep qualities, but often require manual switch between different activity modes. In other words, in order to benefit from those solutions, people have to initiatively use those devices or mobiles apps to track their data if they can "remember" or be bothered. The work presented in [14, 15] are interesting in a sense that automatic sleep detections are implemented using smartphones with only very little behaviour change required from users (mainly at feedback stages).

**Quantified-Self, Personal Informatics and Big Data**
QS is also defined as personal analytics and personal informatics. Li [1] proposed the term personal informatics with a stage-based model that is composed of five stages (preparation, collection, integration, reflection, and action), and identified issues people may have in each stage. In his work, he indicated the barrels for both knowledge discovery and knowledge maintenance. This work shows a clear and useful guidance for people to work with QS in a systematic manner. As stated by Swan [20, 2], one of the crucial conceptual that comes with quantified-self is that with all the user lifestyle data available, future healthcare system will not just is a patient's treatment in a personalized $n = 1$ manner. But the patient, really a participant, or simply a person, becomes the nexus of action-taking and empowerment. The individual, now through quantified self-tracking and other low-cost newly-available tools, has the ability to understand his or her own patterns and baseline measures, and obtain early warnings as to when there is variance and what to do about this.

Swan [2] also has listed several challenges and opportunities that QS brings to big data community including data storage, data integration and data analysis. Typically a single integrated sensor platform is not available for monitoring participants, say combining sleeping quality with daily activity data (say, exercise amount, food intake, stress level and etc.) but instead the challenges of time alignment, normalised sampling rates and handling missing or error-some data have to be addressed directly as presented by Roantree [21]. Bentley [22] and his colleagues have built a health mashup system identify connections that are significant over time between weight, sleep, step count, calendar data, location, weather, pain, food intake, and mood. These significant observations are displayed in a mobile application using natural language. This work supports an increased self-understanding that lead to focused behaviour changes.

**Engaging People for Quantified Self**
Social network based service is the mainstream method that is applied to encourage user engagement. Industry wide, almost all the wearable device providers have their own social network community with hopes that users will compare their activities results with others, thus improving the user engagement [11] [12] [13]. Kamal [23] and his colleagues used social network model to track the health behaviour change and engage users. In their prototype system, users are required to provide various life aspect data including mood, entertainment, food and etc. through a web portal based manual logging system. As discussed earlier, this sort of system requires very strong desire of people to engage and is unlikely to be put into practice in a large scale.

Another method is based on augmented reality (often in forms of gaming). Fitness devices and applications are integrated with games running on smartphones. People have to finish particular tasks in real world (e.g. jogging for a mile, walking to some places in the city, 40 push-ups in a minute) to proceed further in the game [24] [25] [26]. Our view on this sort work is that although they are interesting and may be attractive to some of the users (gamers), the "one-size-fits-all" model should be changed to personalised ones that are adaptive to different individual's needs.

Reminding services are also widely adopted by QS applications. Users can set goals, active level or time thresholds as alarms for triggering the reminding services [11] [13] [12], again, if they have the intention to do so. Recently, more interesting advances of reminding services have been developed. Sleep as android [27], an android smartphone based sleep detecting app can wake people up from sleep without requiring them to setup the alarm clock beforehand. The how it works is that it detects people's sleeping cycle silently and then use the found pattern to decide when the best time to wake people up is. Although, after 3 weeks testing, we found it is still quite immature, it does present an excellent idea-smart devices can play a proactive role in people's life rather than a reactive one that we usually see.

**Paternalism, Libertarian Paternalism and Nudge**
As defined by Gerald back in 1972, paternalism is behaviour by an organization or state which limits some person or group's liberty or autonomy for what is presumed to be that person's or group's own good. Paternalism can also imply that the behaviour is against or regardless of the will of a person, or also that the behaviour expresses an attitude of superiority [7]. Thaler&Sunstein in early 2000's coined the term "libertarian paternalism" [28] meaning that it is both possible and legitimate for private and public institutions to affect behaviour while also respecting freedom of choice. As further explained by the authors, libertarian paternalism is a relatively weak, soft, and nonintrusive type of paternalism because choices are not blocked, fenced off, or significantly burdened. As an implementation of libertarian paternalism, the term "nudge" was proposed-"*A nudge is any aspect of the choice architecture that alters people's behaviour in a predictable way without forbidding any options or significantly changing their economic incentives*". Although the nudge concept was initially stemmed from behaviour economics, work reported in [29] [30] shows that nudging approach may also help generate positive impacts in public health domain.

**PROBLEM ANALYSIS AND DESIGN RATIONALE**
From the current literatures, we can see that most works from the QS domain are heavily focused on the data collection and knowledge discovery stages. In fact, some of the discussion has been simplified in these terms: "having greater awareness of one's behaviour is sufficient to determine a better behaviour change for that individual", which obviously is untrue as the mechanisms that govern people's behaviour change are complex [31]. It is simply not enough to logging and reviewing our calorie intake by a smartphone app (knowledge discovery) to make people want to do more gym exercise or control how much they eat constantly. Also, knowing having had inadequate amount of sleep in the past days doesn't drive people to sleep earlier or more regularly. People often struggle with many constraints such as time, social responsibilities and much more, which is likely to continue even after we are fully aware of the potential problems. In addition, researches have clearly shown how often people do not make decisions on rational basis, but on irrational thinking, such as heuristics and rules of thumb [28] [32]. It's quite clear that simple presentation of data and the awareness of people's own condition are not enough to motivate people to modify their habits.

Therefore, we'd rather take an indirect approach (nudge) with which people do ot need to spend time and energy on understanding themselves (if they don't want to) or consciously change their behaviour and habits, but only need to follow advices that they get and act accordingly for gradual behaviour change. In addition to tracking and presenting people lifestyle data, the goal of our research is to investigate methods that could effectively engage users for a long time and nudge changes towards better lifestyle, with the purpose to achieve ambitious goals such as making people happier, more social, or more efficient in the achievement of their objectives. The fundamental design rationale behind our work is that we try to make QS applications "*effortless*" for people to use or in other words, people should not be aware of they are using a QS application or device but only take the given advices and behaves reactively. In this way, a QS application or device become a proactive service which constantly advises people what to do. This proposed idea is actually in line with some of the existing works that collect users' data unobtrusively. (Such works aim at making data collection effortless). However, to make our idea a reality, three primary questions need to be answered.

***What are the contents of advices that people would find useful and is willing to accept?*** If we think about how human solve problems, the rationale process is in the following order: 1> discover the problem, 2> Understand the problem 3> find solutions for the problem and 4> take actions based on the solutions. This process is exactly what many QS methods are following: using data tracking to help discover problems; using data visualisation and statistical methods to help with better understanding; using social communities or specialists' experience to find solutions and no comprehensive solution for step 4 which is actually the most important step. We argue that general public don't need to be initiatively involved in the first three steps if they want or have time/energy to do so. They need to know what actions to take and how to proceed as what they care about are the final results (e.g. healthier, happier). Therefore, unlike existing QS knowledge representation methods (charting, statistical summaries and etc.), our work aims at providing advices which people can take direct actions upon. For example, instead of showing people how much sleep they had in the past week and hope they will then change their behaviour patterns, direct advices such as "maybe you should continue reading your favourite book in bed now?" maybe a more effective action plan for better sleep. However, in order to provide such useful advices, it crucial to carry the first three steps properly which is illustrated in the following sections.

***When is the best moment to send people advices so those advices could have better chances to be put into actions?*** Conventional reminding services run on a reactive model with which people have to manually setup alarms in advance; get reminded when the alarms are triggered; and take actions accordingly. With such a model, people's actions would normally align with the contents of reminders as they set those themselves. However the proposed "effortless" service works in a proactive way. It should send people advices at the right time when it believes that people need them and have good chance to follow. Such advices cannot be pre-set manually, as they are based on dynamic contexts around people. For instance, it would be irrational to send people advices like "drink a cup of milk now" with the intention for better sleep at noon time if the targeted user has never slept at that time or does it very occasionally or he/she is actually shopping in a supermarket. How to find the right time is a challenge.

***What are the bestpresentations for those advices, which will help engage people better?*** The presentation for the advices is also a very important factor that affects how people react to the given advices. For presentation, we do not refer it to pop up dialogs, text messages or sound alarms as those mechanisms are designed for attracting attentions from people rather than pushing them into actions. What we are really interested in is how to improve the acceptance of the advices sent. For example, sending users a nice picture with a cup of milk and aloud ringtone to their smartphones only reminds them what should be done. How about motivate them to take a picture of the milk that they are going to drink through some game playing and award this behaviour?

Our research aims at seeking answers for these questions. It's crucial to understand 1> what are people doing regularly and at what time? 2> what are potential problems of their life (mainly health related)? 3> what are the interruptible time slots during their daily life?

## RESEARCH DESIGN

Understanding people's lifestyle is a giant topic. In order to control the research scope, we use good sleep quality maintenance as an illustration example. Since as a starting point, it's easier to collect feature data for sleeping pattern and for most of people, sleeping shows more regular patterns than other life activities such as mood. While we were carrying the study, all the methods, system designs and prototyping followed the "*effortless*" principle (Users are not supposed to input a lot nor are forced to give feedbacks explicitly). We used Android smartphones as the main hardware for carrying out all the experiments as they are easy to get and provide richer means for user interactions than those wearable devices in the markets. The proposed work consists several parts. Figure 1 shows the basic workflow of how this work was implemented as well as how the following sections in the paper are structured.

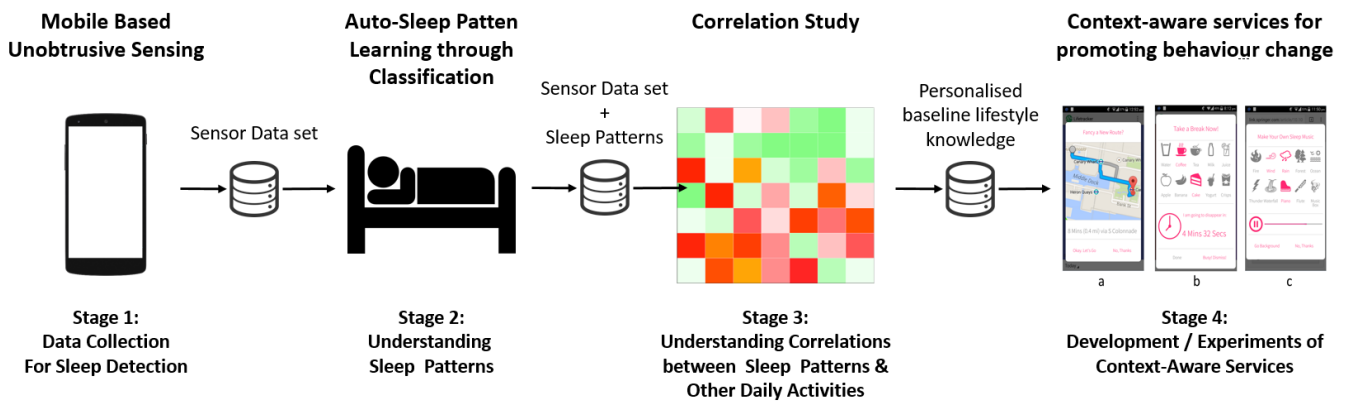

Figure 1: Workflow on how this research is conducted

To test the effectiveness of the proposed research, we conducted a small-scale deployment using 20 subjects (Age groups ranging from 20 to 40. 10 of them are university students and the rests are professionals) for two months (covering all stages shown in Figure 1). They were recruited through university internal advertisements and were promised with Google Playstore vouchers. Only subjects who have owned 4.3+ Android phones were allowed to participate the study and none of them has strong desire to change their current lifestyles after initial interviews. During the study, all the subjects were only advised to use their phone regularly without being asked for significant behaviour change (this only applies to the data collection stage, during later stage when they have to decide whether to follow the advices received, they certainly had have changed some of their phone using behaviour). In the later stage of the research study, all subjects were also fully aware that they will be receiving pop up reminders on their phone in an unpredictable manner (what and when) and they all agreed to keep using the phone as usual instead of removing the app if they find it invasive.

While we carry this research work, the assumption that we made is that 1>all subjects constantly carry phones with them throughout the day and use them regularly; 2>their lifestyles could be well reflected by their phone usages. By displaying significant observations about a person's wellbeing on a smartphone, we aimed to encourage reflection on wellbeing on other devices or combinations of those too.

In the following sections, we will give detailed explanation on methods for each of the steps that we take towards our goal and their associated experiment results.

## STAGE 1: DATA COLLECTION

In order to find as much information as possible about a user's lifestyle, we tried to get all sorts of data that can be collected from a smartphone for later analysis (see Table 1):

| Motivations | Data Sources | Sampling Rate |
|---|---|---|
| **User Location** | GPS, Network Provider, WIFI hot points. | Every 5 minutes |
| **User Places** | Inferred from GPS User Location Data and Google Place Services | Every 5 minutes |
| **Weather** | Inferred from User Location Data and Open Weather Services | Every 5 minutes |
| **Movement** | Accelerometer Sensor | Aggregated Every 5 minutes |
| **Steps** | Inferred from Movement Data | Calculated Every 5 minutes |
| **Walking Time** | Inferred from User Location Data and Steps Data | Calculation Triggered while needed |
| **Running Time** | Inferred from User Location Data and Steps Data | Calculation Triggered while needed |
| **Environment** | Ambient Temperature Sensor, Humidity Sensor, Light Sensor, Microphone | Aggregated Every 5 minutes |
| **Screen On/Off Periods** | Background Services | Every 10 seconds |
| **App Usage** | Background Services | Every 10 seconds |
| **App Type** | Inferred from App Usage Data and Our Own App Type Repository | Every 10 seconds |

**Table 1: Data Collection from an Android Smartphone**

An android app (Lifetracker as shown in Figure 2) was developed and loaded into the smartphone used by each subject. It continuously collects and records the entire feature data listed in Table 1. Collected data are synchronised onto our cloud service [33, 34] continuously every 30 minutes. The system was designed in this way because we intended to carry out all data analysis tasks on the server side given the limited power of mobile devices and battery life.

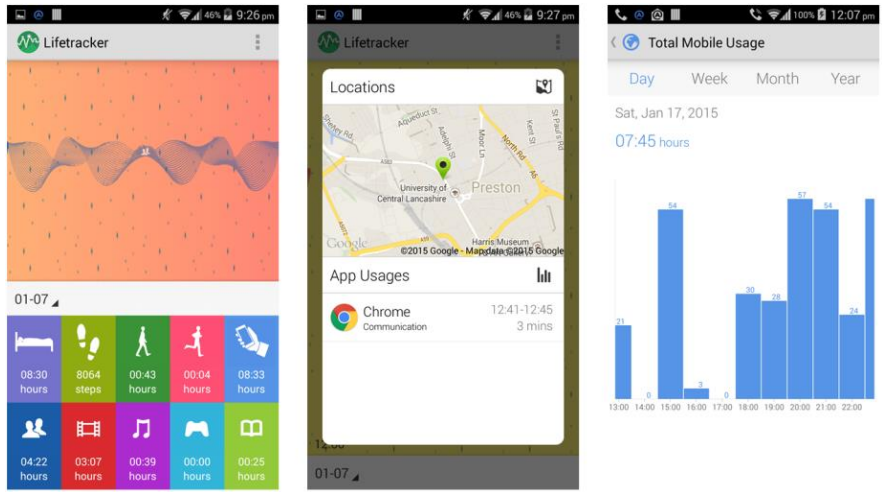

**Figure 2: Lifetracker on Android**

All the subjects are also able to view their data via a web portal as shown in Figure 3. The aim for this portal is different from the normal QS charting tools, as it's provided only for verification purposes rather than expecting users to use it to learn significant amount of knowledge about themselves.

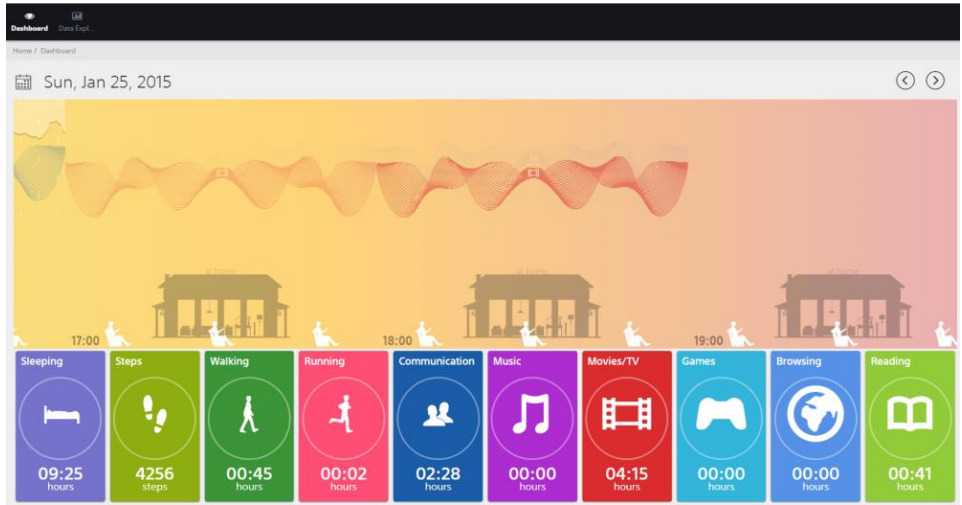
**Figure 3: Web Portal for Verification Purposes**

**STAGE 2: UNDERSTANDING SLEEP PATTERNS**
In order to provide useful advices at the right time, we first need to understand users' sleeping patterns including bedding time, wake up time, sleep during and activities that users perform before sleep. There have been several remarkable works on unobtrusive sleeping detection using smartphones [14] [15]. We have chosen to use the similar approach that's reported from [15] with some enhancements as explained in this section. It should be noted that for our work, we don't consider sleep qualities but only bedding time, wakeup time and sleep durations, as we believe sleep quality factors such as deep sleep, RMD sleep are difficult to acquire accurately without heave user involvement. Even if users do get involved, the results from such are quite subjective for each individual. As a consequence, this would bring in noises to this study.

**Sleep Detection and Classification Method**
To identify different sleep contexts, we've chosen to use the following features (20 features, see Table 2) from a subset of the data collected.

| Modality | Feature Variables |
| --- | --- |
| **Movement** | (Min, Avg, Max, Std) |
| **Noise Level** | (Min, Avg, Max, Std) |
| **Lightness Level** | (Min, Avg, Max, Std) |
| **Screen on/Off periods** | (Min, Avg, Max, Std) |
| **User Locations** | (Latitude, Longitude) |
| **Sleep Time** | (Previous Bedding Time, Previous Wake time) |

**Table 2 Selected Features for Sleep Detection**

The above table shows the best features that we have experimented with. The feature selection process started with a larger modality space which also includes battery states (charging, non-charging, plugged-in, un-plugged), phone powered on/off time. However, including those features didn't seem to give us significant performance boost. After looking into the feature space more carefully, we realised that the phone on/off time is actually correlated with screen on/off period that we choose. Also the battery states vary largely for different brands of android phones. For example, subjects who use HTC phones reported that their batteries usually last for two days on a single charge. However, Samsung Galaxy users do have to charge it every day or every half a day. Another interesting finding was the use of user location as a feature, which, as far as we are aware, has not been reported in others' work. As 10 of our subjects are university students, they show very inconsistent sleeping patterns during the period of this study, which led to poor classification performance. One of the significant issues is that they slept in different places (working on a group project late; Friday clubbing; visiting families during weekends are the root causes). Once the places where they sleep changed, the whole contexts followed such changes dramatically. This is especially true for features such as lightness level, movement as well as noise level. Based on such an observation, during the data collection stage, we tagged all the feature data with user locations and later on, during the training process, we trained classification models only using data from the same locations (in our experiments, places in radius within 100 metres range are regarded as a same place). Then, the model that has the same location tags with the time series data is chosen for new classifications. With our approach for each subject, there exist several classification models with different parameter values. Also, the later evaluation results were based the average of overall performance of all models for each individual.

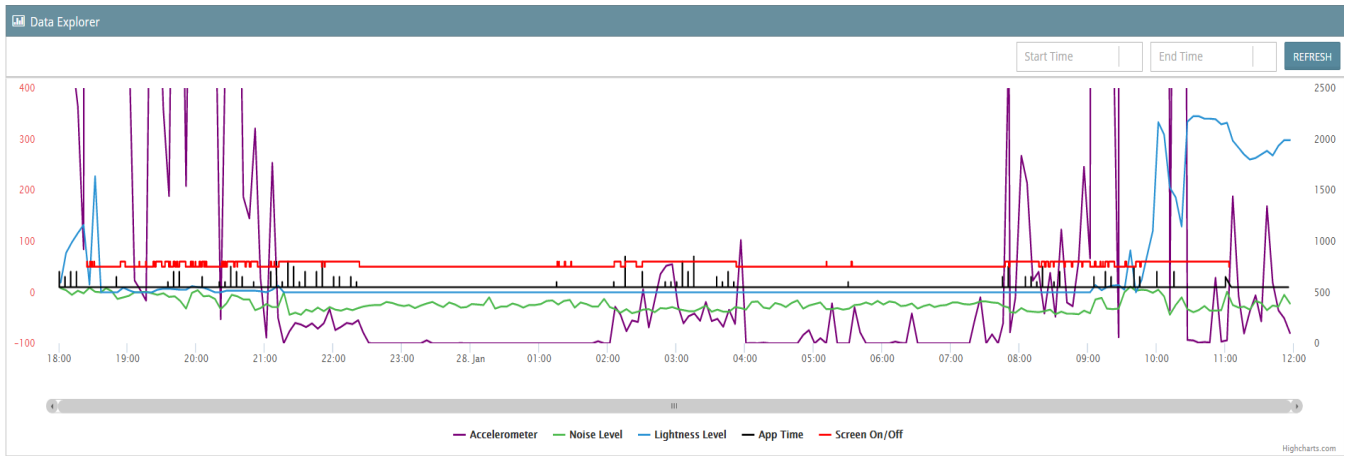

**Figure 4: Time Series Data for User Activities**

The time series data for users' activities are displayed in Figure 4 (also as a part of our web portal). To train the classification models, we divided them (from 18:00pm last day to 18:00pm current day) into non-overlapped 10 minutes windows. The window size is determined through experiments using different classification models. We tried 4 different window sizes (10 minutes, 15 minutes, 20 minutes and 30 minutes). The 10 minutes window performed the best over the other sizes on all the models that were tried. For a new classification, after all the windows are classified, we merge them into bigger chunks in order to gain feedbacks from the participants. The merging principle is that if the time between two positive windows (classified as sleeping periods) is less than 30 minutes, they will be merged. The time width was decided based on the principle that more than 30 minutes could be considered as not-sleep or a sleep disturbance. The starting time of the first merged chunk on the timeline is marked as bedding time and the end time of the last merged chunk is the wakeup time. Sleep duration is the summed length of all merged chunks. For this research, the classification models are trained on individual's basis. We didn't intend to train a general model that works for all users. This is due to the different hardware used on users' phones. The values of features such as accelerometer, ambient light sensors do vary largely on different hardware.

**Gathering Feedbacks**

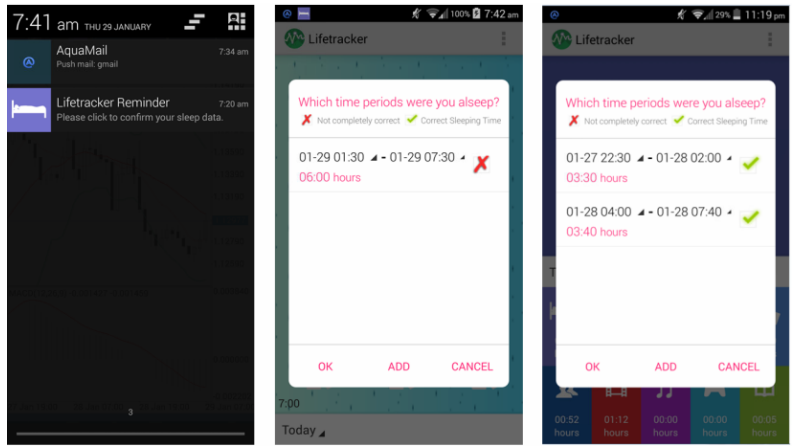

**Figure 5: Gathering Feedbacks for Sleep Contexts**

Gathering feedbacks for model training has been considered as a main issue for works like ours in the literature. It's difficult to keep the participants initiatively log their sleep period over a long time. It's very common to see that people are not able to tell when they go to sleep. Based on our observations, many people just fall in sleep while they are reading book or watching movies. Also, if they forget to log their sleep on a day, it's usually not possible for them to recall that information afterwards unless clear clues are given. We, therefore, designed our feedback mechanism in a proactive manner.

A background service is implemented and is constantly running on users' phones (shown in Figure 5). Every half an hour, it checks new feature data, classifies and merges them. Once it detects a sleep period, it generates a system message to remind users about it for confirmation. All what users need to do is to click the listed items in a popup dialog to confirm the detected sleeping duration if they are correct or change them into correct ones through selecting time if they are incorrect to provide ground truth. If a reminding message is overlooked, a new one will be generated in the next round until user confirms their sleep time. This design largely reduced the noises in the data that we often see from others' work.

Results

| Features | Classifiers' Performance (Positive Classification/Is Sleeping) | | | | | | | | | | | | | | | | | | | | |
|---|---|---|---|---|---|---|---|---|---|---|---|---|---|---|---|---|---|---|---|---|---|
| | Bagging (Fast Decision Tree Leaner) | | | | | Bayesian Network (Simple Estimator) | | | | | KNN (K=6) | | | | | Random Forest | | | | |
| | % | Pre. | Rec. | F. | RRSE(%) | % | Pre. | Rec. | F. | RRSE(%) | % | Pre. | Rec. | F. | RRSE(%) | % | Pre. | Rec. | F. | RRSE(%) |
| $M_1$: Accelerometer {mean, min, max, var} | 81.89 | 0.621 | 0.745 | 0.676 | 79.43 | 80.86 | 0.591 | 0.834 | 0.691 | 91.88 | 81.26 | 0.603 | 0.781 | 0.680 | 81.24 | 80.44 | 0.591 | 0.752 | 0.662 | 83.77 |
| $M_2$: Ambient Light {mean, min, max,var} | 81.01 | 0.457 | 0.477 | 0.465 | 79.83 | 81.08 | 0.613 | 0.833 | 0.694 | 91.13 | 80.575 | 0.615 | 0.789 | 0.676 | 79.32 | 80.36 | 0.612 | 0.786 | 0.672 | 80.24 |
| $M_3$: Screen Status {mean, min, max,var} | 74.23 | 0 | 0 | 0 | 98.26 | 68.08 | 0.282 | 0.635 | 0.390 | 98.21 | 74.231 | 0 | 0 | 0 | 91.43 | 74.16 | 0 | 0 | 0 | 91.56 |
| $M_4$: User Location {latitude ,longitude} | 75.38 | 0.544 | 0.058 | 0.103 | 96.28 | 75.29 | 0.561 | 0.048 | 0.088 | 95.74 | 75.481 | 0.517 | 0.078 | 0.128 | 94.70 | 75.57 | 0.525 | 0.078 | 0.129 | 94.54 |
| $M_5$: Time {start, end in hour} | 89.717 | 0.782 | 0.829 | 0.805 | 63.09 | 89.49 | 0.756 | 0.868 | 0.808 | 66.54 | 89.79 | 0.792 | 0.814 | 0.802 | 61.936 | 89.79 | 0.793 | 0.812 | 0.802 | 61.91 |
| $M_6$: Ambient Noise {mean, min, max,var} | 73.13 | 0.251 | 0.094 | 0.127 | 93.22 | 70.06 | 0.453 | 0.718 | 0.545 | 102.25 | 71.131 | 0.246 | 0.174 | 0.194 | 94.60 | 71.57 | 0.282 | 0.137 | 0.171 | 95.85 |
| {$M_1$:$M_6$} | 95.302 | 0.902 | 0.918 | 0.910 | 42.82 | 92.70 | 0.811 | 0.934 | 0.868 | 64.67 | 92.681 | 0.824 | 0.910 | 0.865 | 51.58 | 95.48 | 0.907 | 0.919 | 0.913 | 41.33 |

**Figure 6: Classification Performance for 10 minutes Sleeping Windows (Bold shows the best performer for each of the modalities**

We tried several classification models in order to achieve the best performance. To train those models, we used 25 days data and performed some data pre-processing tasks. This ends up with 3505 10-minutes window instances for each subject. To evaluate those models, 10-fold cross-validation was applied. The results are shown in Figure 6. We can see that none of single feature gave acceptable results apart from the "$M_5$: sleeping time" feature which has to be manually supplied by the participants. However, while putting them together, the performance boosts dramatically. The best result is achieved via using Random Forest classifier, which shows an average of 95.48% accuracy on classifying sleeping windows with F-value at 0.91. The results are similar to the work presented in [15] (ours is slightly better, but we understand this is due to the extra "user location" feature introduced. Also, the selected subjects' behaviour may also have impacts on the results) and provide a solid ground for carrying later steps of our work.

We also investigated how fast we will be able to approach the best results. This is an important factor that would affect how well the service will be accepted by end users, as real users don't have enough patience to keep logging their data for a long time (the fact is users who have less initiative normally give up logging quickly if they don't see the automatic detection working in a short term).

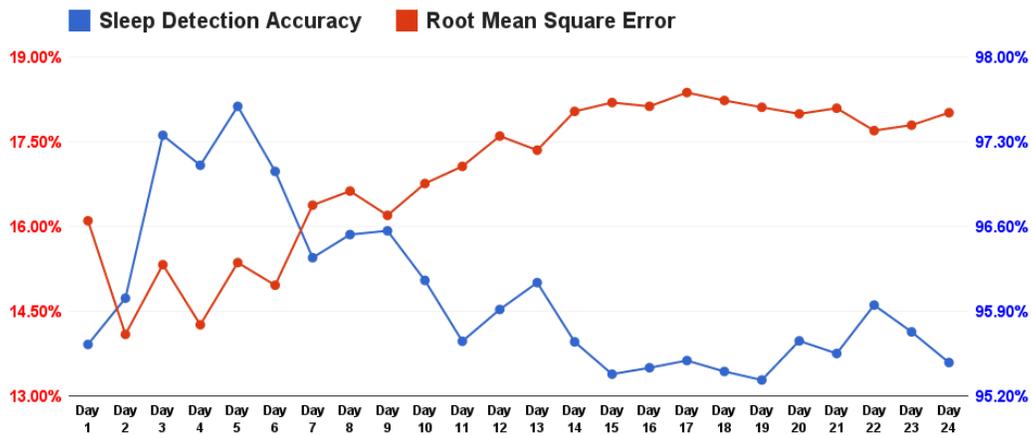

**Figure 7: Sleep Detection Accuracy and RMSE change over 24 days.**

Figure 7 shows the sleep detection accuracy and RMSE change over 24 days, from which we can see the performance of the model did decrease along with the time till day 15 and got stabilised afterwards. Although it performs well for the first 6 days (97.59% accuracy highest and 14.09% RMSE lowest), we interpret it as over-fitted due to insufficient amount of data collected. However, we don't consider this as a negative aspect from user experiences' point of view. The average starting accuracy after one day is as high as 94.63% with RMSE at 16.10%, which means users are very likely to see accurate activity detection without be required to supply large manual inputs at very early stages. This would encourage them to keep using the service "*effortlessly*". Also, the performance drop later is not significant enough to discourage them as they won't even notice (the lowest average accuracy is 95.38% that is still high enough to avoid user inputs).

**STAGE 3: UNDERSTANDING POTENTIAL PROBLEMS AND CORRELATIONS BETWEEN SLEEP PATTERNS AND OTHER ACTIVITIES**

After we gained understanding of people's sleep patterns, we then tried to find out what are the potential problems that people experience and what are the correlations between those patterns (either good or bad) and other activities/habits from their daily life. As such study is on individual bases, we choose subject 8, who is a professional in financial industry, as an illustration example.

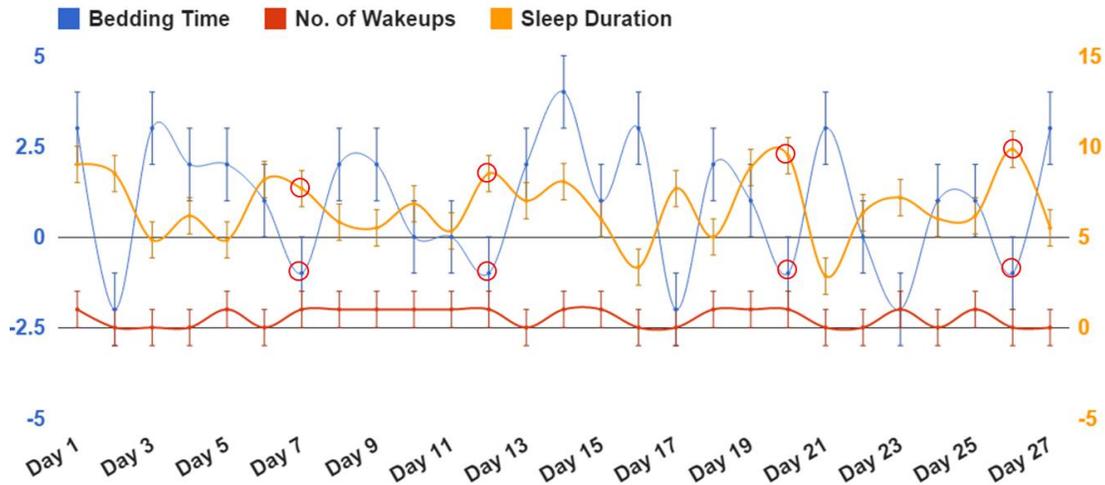

Figure 8: Sleep Patterns of Subject 8 in 27 Days

From the above figure, we can see that subject 8 was experiencing quite irregular sleep patterns. Her bedding time varied 2.5 hours across the study period with minimum sleep duration of 3.7 hours. Also she often woke up during her sleep. Lifestyle like this (disrupted sleeps) will potentially lead to more serious healthy issue in a long run for her [35].

We used as much context information as possible for the correlation study. Parameters considered are shown in Table 3.

| Parameters | Period | Units |
|---|---|---|
| Daily Steps | From the last wakeup time to the current bedding time | Number of Steps |
| Daily Walking Time | From the last wakeup time to the current bedding time | Minutes |
| Daily Running Time | From the last wakeup time to the current bedding time | Minutes |
| Daily Communication Time | From the last wakeup time to the current bedding time | Minutes |
| Daily Video Watching Time | From the last wakeup time to the current bedding time | Minutes |
| Daily Music Listening Time | From the last wakeup time to the current bedding time | Minutes |
| Daily Reading Time | From the last wakeup time to the current bedding time | Minutes |
| Daily Gaming Playing Time | From the last wakeup time to the current bedding time | Minutes |
| Lightness Level | 2 hours before the current bedding time | Minutes |
| Noise Level | 2 hours before the current bedding time | minutes |
| Lists of Apps Used | 2 hours before the current bedding time | minutes |
| Bedding Time | N/A | Clock Time |
| Wake up Time | N/A | Clock Time |
| Sleep Durations | N/A | Hours |
| Number of Wakeups | Between Bedding and Wakeup time | Number |

Table 3: Parameters Used for Correlation Study

It should be noted that for the correlation study all used data are time stamped before the current bedding time except those sleep pattern data. The rationale behind this choice is that correlation study normally doesn't show enough information about causalities which are in fact what we need for this work. (We want to find activities/causes that lead to bad sleep patterns but not the other way around). In addition, we are fully aware that bad sleep patterns could be caused by many factors such as mood, alcohol intake amount, stress level from work, social relation problems and etc., which are not directly measureable (at least unobtrusively) at the moment. However, they can be incorporated into our framework while technologies advance further. For our study, we stick with the "effortless" principle for now and didn't ask user to input those information manually.

**Results**

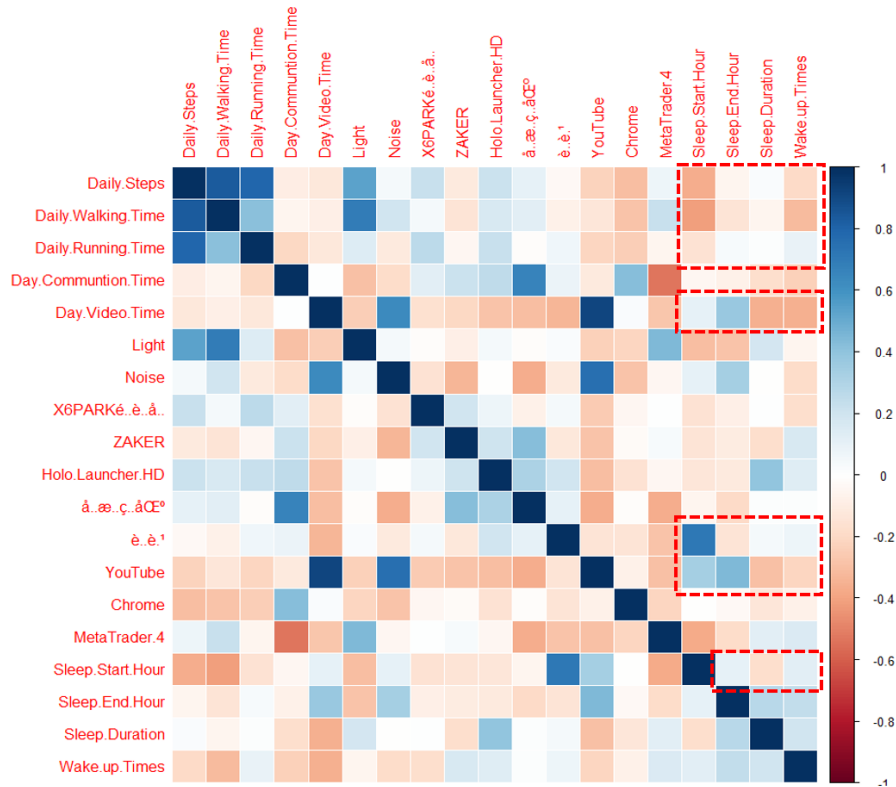

**Figure 9: Correlations between Sleep Patterns and Other Activities/Contexts of Subject 8**

The heat map in Figure 9 shows some interesting correlations about this subject as highlighted in the dashed boxes. We can see that for this particular participant, daily steps/walking time/running time show noticeable negative correlations with bedding time and number of wakeup times during sleep. This likely tells that more daily exercises could be helpful with her sleep pattern. Total daily video watching time also has negative correlations with number of wakeup time. She is less likely to wake up during night if she watches a lot of videos during the day. However, video watching time does have negative correlations with sleep duration, which indicates the more videos she watches, the less sleep she gets. Moreover, two apps she uses have significant correlations with her bedding time. Finally, sleep earlier (but only between 11PM and 12AM as cycled in Figure 8) does help with her sleep pattern as it both increases sleep duration and decreases number of wakeup times. Besides analysing activity correlations, the service also calculates the best average steps amount, bedding time, usage time for each apps that users used before sleep, (by "best", we refer to values of those parameters that correlate to longer sleep duration and less wakeup times. These values are used as guide lines for our work in later steps).

**STAGE 4: DEVELOPING CONTEXT-AWARE SERVICES FOR PROMOTING BEHAVIOUR CHANGE**
Based on the analysis results from the previous study, we continue on generating personalised advices for promoting people's behaviour change using context-aware services. As discussed earlier, the key challenge is how to engage users better when they receive those advices and how we can possibly evaluate whether the advices are accepted by the users without explicitly asking for feedbacks. We designed and implemented three context-aware recommending services that are aiming at: 1> improving daily amount of steps; 2> reducing continuous screen time/app usages and 3> helping people relax before/during their sleep.

In the follow sections, we explain in details how each of the services works, the experiments that we designed and experiment results.

**Service One: Nudging for More Steps**
The main approach that we designed to nudging people for taking more steps is to remind them using alternative paths for their reoccurring trajectories (e.g. from home to tube station; from tube station to work and etc.). In order to automatically detect the reoccurring trajectories without requiring too much user inputs, we first need to identify all their walking routes on a single day. Figure 10 shows the basic idea.

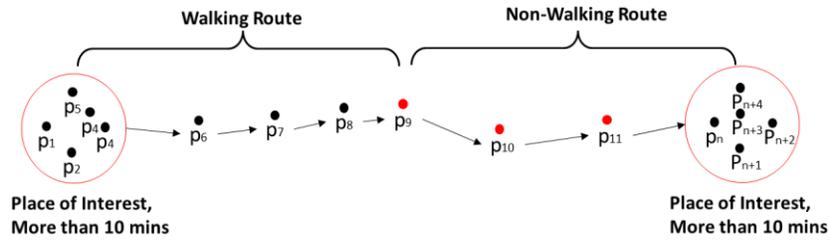

**Figure 10: Walking Routes Detection**

For each day (except Saturday and Sunday as we found walking routes on these two days show large variations), we first sort all the user's location data between his/her last wakeup time and the current bedding time. Then we cluster all location points that are connected sequentially (with regards to the location timestamps) and are within 10 metres distance. For each cluster, we calculate their centrals (using KMeans); label the centrals as points of interest and use them as "break points" to separate all location data from that day into several segments. At the last, for each segment, we filter out the consecutive data points that have higher distances than a pre-set threshold (for our work, we use 150 metres, as our data collection frequency is 1 minute).

To learn reoccurring trajectories, we used all working routes from two weeks and compare the similarities between them iteratively. To measure the similarity between two walking routes, we've applied a much simpler method than those from the literatures [36, 37, 38]. For a pair of routes that needs to be compared, Google places service [39] is queried to obtain street name for each of the location points (house numbers and postcodes are removed and same street names for one route are discarded). The result street names are then concatenated into a larger string in the same order as the location points for each walking route. Finally, we calculate the edit distance between the two final strings and if the value is less than a pre-set threshold, the two walking routes are regarded as a same trajectory (see Figure 11).

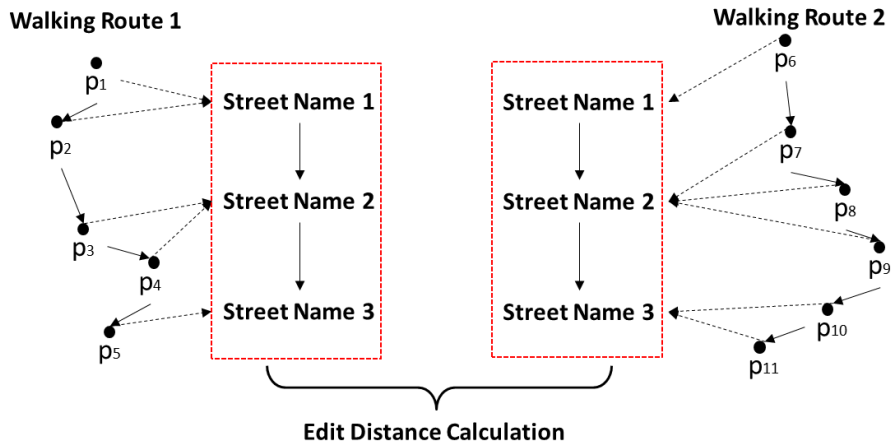

**Figure 11: Route Similarity Detection**

All the reoccurring trajectories found are tagged with their starting location, end location, starting clock time, end clock time, average steps taken in between, average walking speed, map distance (obtained from Google navigation service) as well as all street names. Given these information, alternative routes (using Google Map API) that connect the same start location and end location for a reoccurring trajectory are retrieved and stored locally, so are the most probable week days and clock time (in hour) for it. A background service was implemented and runs every 15 minutes. It searches for all reoccurring trajectories for the day as well as the clock time for the immediate forthcoming one. If it finds one, it fetches the associated alternative routes and calculates steps amount for each based on the average walking speed learnt from the existing trajectory. The alternative route that costs more steps will be advised to the user. It should be noted that as a user may have multiple reoccurring trajectories on a single day, we use the one that had the "best" step amount from our correlation study as a guideline. Differences between the "best" route and sum of steps for each of the potential walking routes are calculated for selecting which alternative route should be suggested. We didn't use the route that has the largest differences as it may has higher chance to be rejected by the end user. If no alternative route if found or one alternative route has been continuously rejected for more than 3 times, no advice will be generated for that trajectory anymore.

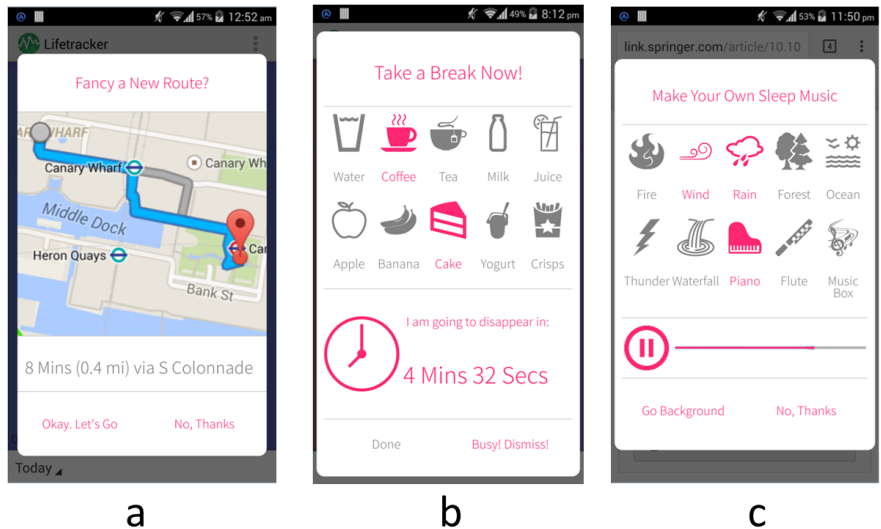

**Figure 12: Nudging Services for Encouraging More steps, Less App Use Time and Relaxing before Bedding**

After an advice is generated, it is cached on the phone and only got sent when two conditions are met: 1> half an hour before the reoccurring trajectory is about to start; 2> and while the user is using his phone. Whether a user is using the phone is detected in real time by calculating the accelerometer readings, screen on status in a 5 minutes cycle. In this way, we have more confidence that the advice is actually seen by the user. Users in our study only have two choices while they see an advices. They can either click "Okay, Let's go" which means they accepted the advice or "No Thanks" to reject it (as shown in Figure 12.a).

**Results**

To evaluate the performance of the above service, we carried out our experiments over 20 days. The first 10 days were used to calculate reoccurring trajectories and the last 10 days were used to generate advices. In the experiment, we tested how many advices were seen by users and were accepted in the last 10 days against the total reoccurring trajectories that actually took place in the first 10 days. All participants were randomly divided into two groups of equal size of 10. In the first group users' app, the alternative route advices were generated using our proposed approach while for the second group of users, advices were generated randomly before a reoccurring trajectory takes place on a day.

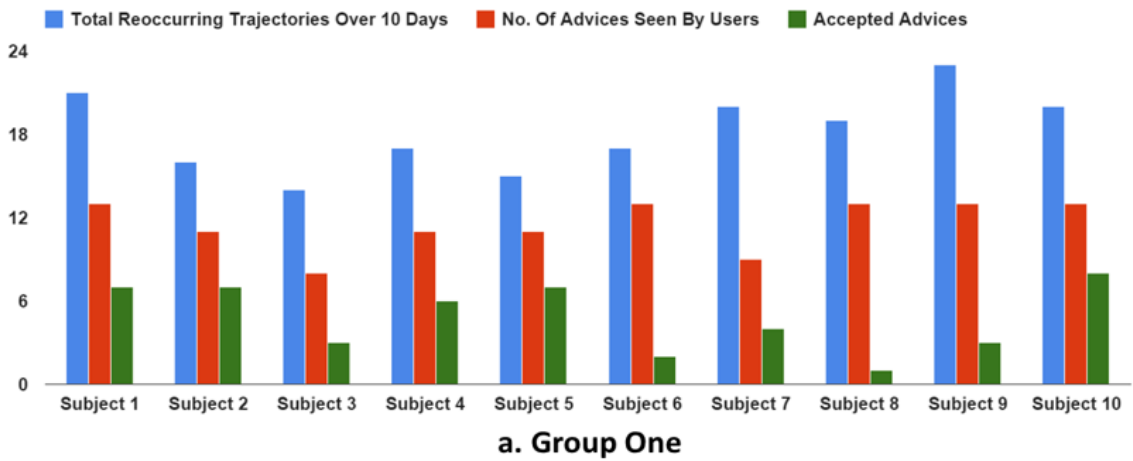

**Figure 13：Total Reoccurring Trajectories VS No. of Alternative Route Advices Generated VS No. Of Accepted Advices.**

The results are shown in Figure 13, from which we can see that with our approach (**Error! Reference source not found.**.a) for group one, 63.19% of alternative advices were seen by users (an advice is only considered as "seen" if it has an user feedback recorded) with acceptance rate at 41.73% . To gather ground truth on whether a user has accepted an advice or not, comparison between the actual walking routes with the suggested walking routes are conducted for each advices.

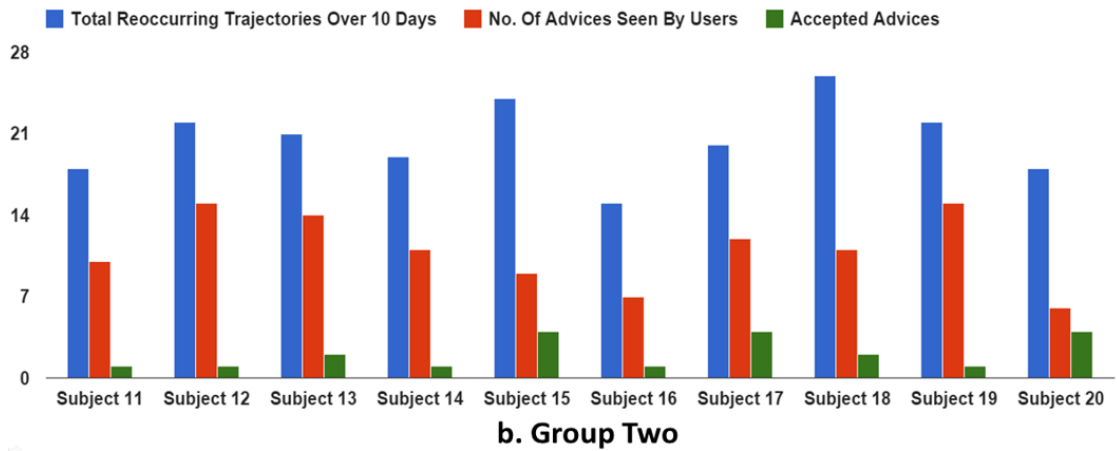

**Figure 14: Total Reoccurring Trajectories VS No. of Alternative Route Advices Generated VS No. Of Accepted Advices.**

For group two, (Figure 14), although similar amount of advices were generated seen by users (53.68%), the acceptance ratio is much lower at 19.09%. This confirms that even we know users activities/behaviours, when to remind them to trigger behaviour changes will affect the results significantly.

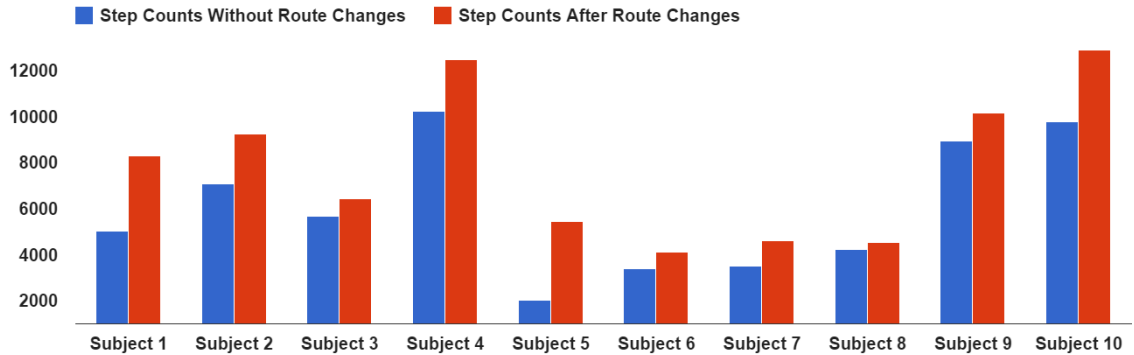

**Figure 15: Step Counts Changes (without Route Change VS with Route Changes)**

There is also an average 3.03% steps increasing for the first group users in 10 days as shown in Figure 15. This amount doesn't seem to be significant at first glance. However, with our approach, if users keep accepting advices in a long run, this number will likely keep going up as the new routes that we advise always have higher step counts than the current one. The trend will continue till it coverages at some points (e.g. the longest path is taking by users, or users choose to use a shorter path constantly).

**Service Two: Nudging For Less App Time Usage**

Although addiction to mobile devices or apps is a well-known problem [40], most of the existing works mainly target at kids' mobile usage control. Very limited effort work can be found for adults to the best of our knowledge. Using password to stop people using their mobile phones or pre-set a limited use time cannot effectively reduce the mobile use time for those who have less initiative. Apps that are designed this way normally end up with being uninstalled as they are very likely to disturb the normal phone usages. Instead, our work for tackling this problem focused on gentle disturbances. As shown in Figure 12.b, we developed a service that sends advices to users to remind them for breaks. The service doesn't explicitly ask users to stop using their phones/apps, but only suggests breaks for snacks or drinks. Moreover, it doesn't need to be pre-set for a regular repeating time or for particular apps. It learns when is the best time to send those advices, which is modelled as another classification problem. Once a user accepts the break reminder, he can also log what he has done during the break. Although the logged data such as water intake, calories are not used for this work, they are reserved for feature studies.

**Results**

| Modality | Feature Variables |
|---|---|
| Week Day | Day of Week |
| App Package Name | App ID |
| Continuous App Usage Time | (Min, Avg, Max, Std) |
| App Start Using Time | Clock Time in Hour |
| Total No. of the Same App Use on that Day | (Min, Avg, Max, Std) |
| Movement (+5/-5 Mins around advice generation time) | (Min, Avg, Max, Std) |

**Table 4: Selected Features for Disturbance Advice Generation**

Random forest is adopted as the classification model using 16 features (see Table 4). Total 732 instances were collected in 10 days (sleeping hours were filtered out) from each subject (group of 10) for training. 10-folds cross validation was applied for evaluating the model. The result shows an average acceptance accuracy at 63.23% (Precision: 0.576, Recall: 0.610, F-Value: 0.59, RMSE: 90.23%).

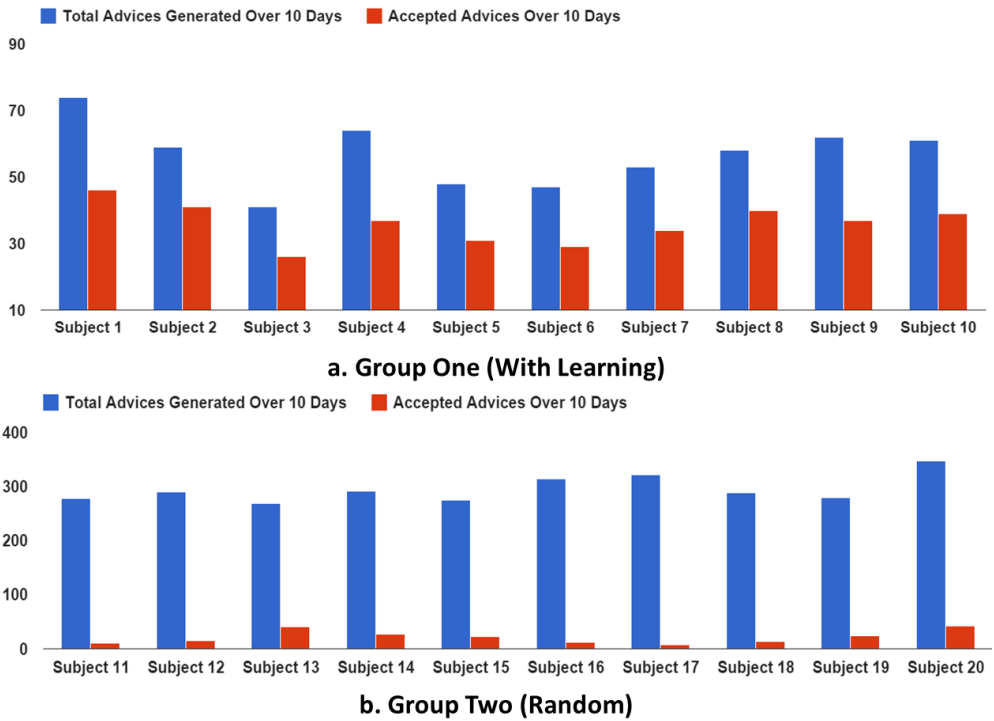

**Figure 16: App Disturbance Advice Acceptance Rate (Leaning VS Random).** It can be noted that subjects in group two have received significantly more advices generated by the app. This is because that the service running at the background only checks if the phone screen is on and if the past accelerator readings are above a particular threshold every 5 minutes in every 30-60 minutes window. This made the result comparison between the two groups difficult. When the experiment was first designed, we were not able to normalise the advice generation frequency effectively as it was not leant yet from Group one. However, for similar future studies, we are able to use the frequency for a better control over group two.

We have to admit this result doesn't look very promising on its own. However, while being applied in later testing stages (another 10 days for testing), it outperformed the results from random advices generation (advices are generated with irregular intervals between 30 minutes and an hour) in terms of the number of accepted advices (See Figure 16, random advice generation only received 7.3% acceptances). In addition, total app usages reduced 10.03% averagely across the 10 subjects in group one with our proposed approach as shown in Figure 17.

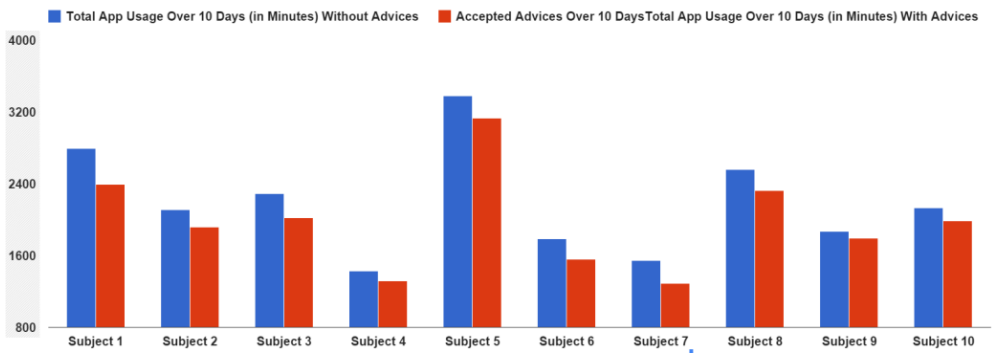

**Figure 17: Total App Use Time Reduction**

We also noticed that, from the experiments, there are no clear correlations between the number of generated advices, the number of accepted advices and the app use time reduction. Further experiments need to be carried out to see the rules behind.

**Service Three: Nudging For Earlier Bedding Time**

The last service that we designed aims at relaxing people before they go to sleep with a "*hope"* that getting them relaxed earlier could make them go to bed earlier and sleep better. The service is designed as a simple game with which people can combine different sound tracks/effects into a melody that they are comfortable with. Sound effects can be added or removed one by one in real time so users are able to hear the change immediately without stopping or restarting the track. The service is ready to be triggered an hour before the "best" bedding time that's learnt from the correlation study for each individual. Similar to the alternative route service, a pop up reminder (see Figure 12.c) appears at the time when a user is using his phone after the reminder is generated. If "go background" is clicked, the sound track keeps playing until it's stopped by users manually or by the background service when it detects that the user is asleep. Also, using the accelerometer and lightness data, the service adjusts the sound volume down little by little. Moreover, the service automatically logs information including the combination of sound effects for every played track; length of played sound tracks; sound track playing start time/end time and how is a play terminated (by user or by service).

**Results**

The experiment for this service ran for another 25 days right after the sleep detection data collection periods. All subjects were aware that this service would be triggered in the evening but were not forced to play with it. Data of participants who had not used the service regularly (less than 10 minutes averagely for each use) are removed from the study. Figure 19&17 shows the change of bedding time variances, before and after service use, where we can see for people who have regularly bedding time, using this service doesn't change the regularity much. However, for the two extreme cases (subject 8 & subject 19), the bedding time variances did reduce to a noticeable level, so are the incensement for their sleep duration.

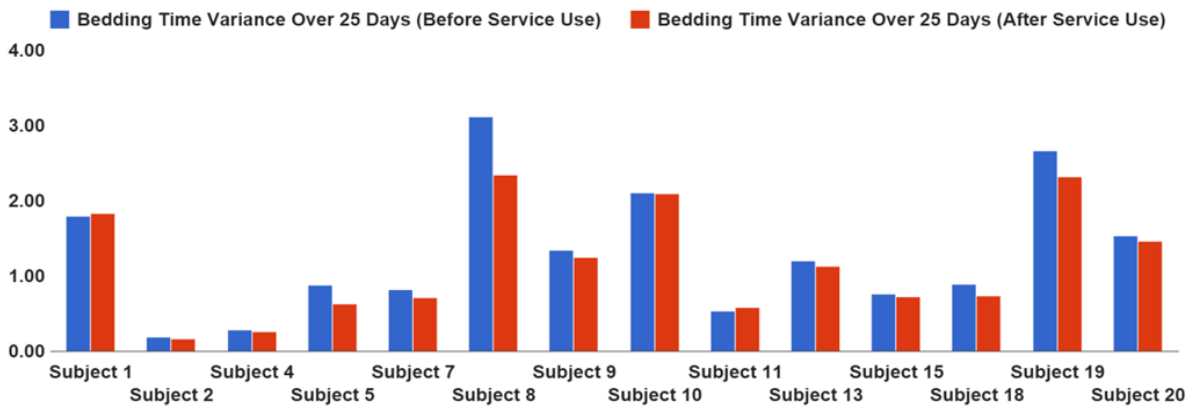

Figure 18.a Bedding Time Variance and Average Sleep Duration Changes Before Service Use

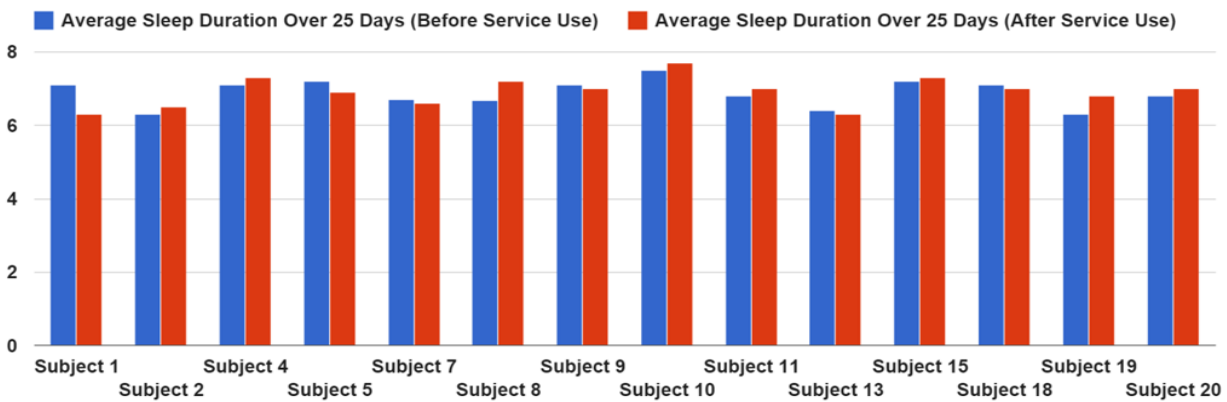

Figure 19.b: Bedding Time Variance and Average Sleep Duration Changes after Service Use)

**DISCUSSIONS**

The work specially targets the group of people who have less disruptable lifestyle but have enough intrinsic motivation to improve their well-being. The proposed framework and associated context-aware services can be considered as extrinsic digital advisers that silently learn people's lifestyle, discover potential unhealthy behaviour and generate advices for gradual behavioural change based on the knowledge learnt, without breaking the main life streamline of users. In short, based on the learning carried in the early stages, the extrinsic digital adviser is able to "guess" user's will at a particular event moment and comes up with an altered action plan which doesn't violate user's will dramatically. From psychological points of view, Based on the self-determination theory (and its second sub-theory: Organismic Integration Theory) [41], our work complies with and contributes to integrated regulation form of extrinsic motivation, which is the most autonomous form.

The main weakness of this work, as far as we are aware, comes from the correlation study in which we assumed that the true causalities of good/bad sleep patterns can be reflected from correlations between people's sleep patterns and other activities. However, this is not always true. For example, although there are correlations between daily steps amount, people's bedding time and sleep duration, increasing step amount doesn't necessarily lead to a better sleep. Large amount of steps may be correlated with other activities that are out of our observations (at least for this study). For instance, walking long distance to workplaces and having stressed work for the whole day are two facts for a day. People may end up with better sleep because they had an exhausted working day, but not because the large amount of steps that they took. Therefore, how to find causalities of peoples' lifestyles properly in an unobtrusive way still remains as a big challenge for now. Development of new types of wearable sensor devices, such as lightweight EEG glass, may provide measurements for good features like mood, stress level and etc. directly.

Another limitation of this work is that the number of participants involved in the study is small and half of them are university students. It's difficult to argue whether the result from this work is reprehensive enough for a general population with larger variety of behaviours. Although the work itself has shown its value and some interesting findings for the purpose of proof of concept, the results of some experiments (especially the last two) need to be explored further with more data. The next step for this work is to publish the android app in Google Play Store to attract more general users. This will help us to collect more data, consolidate the current experiments and elaborate more interesting findings

The last potential problem of this work is it may lead to serious privacy concerns as a large amount of context information about a user can be discovered quite easily including where they live, where they work, their real time movement, means of transportation, their sleep patterns, apps that they use and many more. Although permissions based approaches have been applied on reducing exposure of such information, restrictions like this reduce/eliminate some useful features. This could largely affect the performances of the services developed in this research. However, such an issue exists in almost all human-centred applications or services and is out of the scope of this paper.

**CONCLUSION**

In this paper, we proposed a "nudging" based concept and its associated services, which aim at helping people achieve better lifestyles with less effort. We believe the work presented in this paper has the following contributions to the UbiComp, personal informatics and healthcare communities:

1> The concept of using proactive services for quantified-self applications in order to minimise people's effort.
2> A set of proactive services which learn user behaviour automatically and use learnt knowledge to nudge peoples' activities without requiring too much manual intervention.
3> A highly accurate unobtrusive sleep detection method through using user location based data filtering before model training and a lightweight trajectory planning method which can be used for alternative route advises for many healthcare or wellbeing based applications.

In our future work, we plan to incorporate more features into the existing study. Data for many other features have been collected through using our existing such as water amount, calories intake amount and etc. We also plan to design methods and services that learn people's mood/stress level through unobtrusive interactions.

**REFERENCES**

[1] I. Li, A. K. Dey and J. Forlizzi, "Understanding My Data, Myself: Supporting Self-reflection with Ubicomp Technologies," in *Proceedings of the 13th International Conference on Ubiquitous Computing*, New York, NY, USA, 2011.

[2] M. Swan, "The quantified self: Fundamental disruption in big data science and biological discovery," *Big Data,* vol. 1, no. 2, pp. 85-99, 2013.

[3] I. Li, A. Dey and J. Forlizzi, "A stage-based model of personal informatics systems," in *Proceedings of the SIGCHI Conference on Human Factors in Computing Systems*, 2010.

[4] J. Rooksby, M. Rost, A. Morrison and M. C. Chalmers, "Personal tracking as lived informatics," in *Proceedings of the 32nd annual ACM conference on Human factors in computing systems*, 2014.

[5] J. Kopp, "Self-monitoring: A literature review of research and practice," in *Social Work Research and Abstracts*, 1988.

[6] E. K. Choe, N. B. Lee, B. Lee, W. Pratt and J. A. Kientz, "Understanding quantified-selfers' practices in collecting and exploring personal data," in *Proceedings of the 32nd annual ACM conference on Human factors in computing systems*, 2014.

[7] G. Dworkin, "Paternalism," *the Monist,* pp. 64-84, 1972.

[8] P. C. Shih, K. Han, E. S. Poole, M. B. Rosson and J. M. Carroll, "Use and adoption challenges of wearable activity trackers," *iConference 2015 Proceedings,* 2015.

[9] M. Bazerman and D. A. Moore, "Judgment in managerial decision making," 2012.

[10] *Apple Watch. https://www.apple.com/uk/watch/.*

[11] *FitBit. http://www.fitbit.com.*

[12] *Jawbone. https://jawbone.com/.*

[13] *Nike+. https://secure-nikeplus.nike.com/plus/.*

[14] Z. Chen, M. Lin, F. Chen, N. Lane, G. Cardone, R. Wang, T. Li, Y. Chen, T. Choudhury and A. Campbell, "Unobtrusive sleep monitoring using smartphones," in *Pervasive Computing Technologies for Healthcare (PervasiveHealth), 2013 7th International Conference on*, 2013.

[15] J.-K. Min, A. Doryab, J. Wiese, S. Amini, J. Zimmerman and J. I. Hong, "Toss 'N' Turn: Smartphone As Sleep and Sleep Quality Detector," in *Proceedings of the SIGCHI Conference on Human Factors in Computing Systems*, New York, NY, USA, 2014.

[16] *XinBao Heart Care. http://www.zettasense.co.uk.*

[17] *AliveCor. http://www.alivecor.com/home.*

[18] *Glooko. https://glooko.com/.*

[19] *Withings. http://www.withings.com/.*

[20] M. Swan, "Health 2050: the realization of personalized medicine through crowdsourcing, the Quantified Self, and the participatory biocitizen," *Journal of Personalized Medicine,* vol. 2, no. 3, pp. 93-118, 2012.

[21] M. Roantree, J. Shi, P. Cappellari, M. F. O'Connor, M. Whelan and N. Moyna, "Data transformation and query management in personal health sensor networks," *Journal of Network and Computer Applications,* vol. 35, no. 4, pp. 1191-1202, 2012.

[22] F. Bentley, K. Tollmar, P. Stephenson, L. Levy, B. Jones, S. Robertson, E. Price, R. Catrambone and J. Wilson, "Health Mashups: Presenting Statistical Patterns Between Wellbeing Data and Context in Natural Language to Promote Behavior Change," *ACM Trans. Comput.-Hum. Interact.,* vol. 20, no. 5, pp. 30:1--30:27, #nov# 2013.

[23] N. Kamal, S. Fels and K. Ho, "Online Social Networks for Personal Informatics to Promote Positive Health Behavior," in *Proceedings of Second ACM SIGMM Workshop on Social Media*, New York, NY, USA, 2010.

[24] *Ingress. https://play.google.com/store/apps/details?id=com.nianticproject.ingress\&hl=en_GB.*

[25] *BallStrike. http://www.fit-master.com/.*

[26] *ZombiesRun. https://www.zombiesrungame.com/.*

[27] *Sleep as Android. https://sites.google.com/site/sleepasandroid/.*

[28] R. H. Thaler and C. R. Sunstein, Nudge: Improving decisions about health, wealth, and happiness, Yale University Press, 2008.

[29] P. Basham, "Are nudging and shoving good for public health," *A Democracy Institute Report: http://tinyurl.com/4m6j6m9,* 2010.

[30] S. Vallg{\aa}rda, "Nudge—A new and better way to improve health?," *Health policy,* vol. 104, no. 2, pp. 200-203, 2012.

[31] A. Rapp, "Beyond gamification: Enhancing user engagement through meaningful game elements.," in *FDG*, 2013.

[32] A. Tversky and D. Kahneman, "Judgment under uncertainty: Heuristics and biases," *science,* vol. 185, no. 4157, pp. 1124-1131, 1974.

[33] Y. Li, L. Guo, C. Wu, C.-H. Lee and Y. Guo, "Building a cloud-based platform for personal health sensor data management," in *Biomedical and Health Informatics (BHI), 2014 IEEE-EMBS International Conference on*, 2014.

[34] Y. Li, C. Wu, L. Guo, C.-H. Lee and Y. Guo, "Wiki-Health: A Big Data Platform for Health," *Cloud Computing Applications for Quality Health Care Delivery,* p. 59, 2014.

[35] A. Theadom and M. Cropley, "'This constant being woken up is the worst thing'--experiences of sleep in fibromyalgia syndrome," *Disability and rehabilitation,* vol. 32, no. 23, pp. 1939-1947, 2010.

[36] L. Chen, M. T. {\"O}zsu and V. Oria, "Robust and fast similarity search for moving object trajectories," in *Proceedings of the 2005 ACM SIGMOD international conference on Management of data*, 2005.

[37] S. Dodge, P. Laube and R. Weibel, "Movement similarity assessment using symbolic representation of trajectories," *International Journal of Geographical Information Science,* vol. 26, no. 9, pp. 1563-1588, 2012.

[38] G. Trajcevski, H. Ding, P. Scheuermann, R. Tamassia and D. Vaccaro, "Dynamics-aware similarity of moving objects trajectories," in *Proceedings of the 15th annual ACM international symposium on Advances in geographic information systems*, 2007.

[39] *Google Place Service. https://developers.google.com/places/documentation/.*

[40] M. Choliz, "Mobile phone addiction: a point of issue," *Addiction,* vol. 105, no. 2, pp. 373-374, 2010.

[41] R. M. Ryan and E. L. Deci, "Self-determination theory and the facilitation of intrinsic motivation, social development, and well-being.," *American psychologist,* vol. 55, no. 1, p. 68, 2000.

[42] M. Kay, E. K. Choe, J. Shepherd, B. Greenstein, N. Watson, S. Consolvo and J. A. Kientz, "Lullaby: A Capture \&\#38; Access System for Understanding the Sleep Environment," in *Proceedings of the 2012 ACM Conference on Ubiquitous Computing*, New York, NY, USA, 2012.

[43] M. Morris and F. Guilak, "Mobile Heart Health: Project Highlight," *IEEE Pervasive Computing,* vol. 8, no. 2, pp. 57-61, #apr# 2009.

[44] S. Consolvo, D. W. McDonald, T. Toscos, M. Y. Chen, J. Froehlich, B. Harrison, P. Klasnja, A. LaMarca, L. LeGrand, R. Libby, I. Smith and J. A. Landay, "Activity Sensing in the Wild: A Field Trial of Ubifit Garden," in *Proceedings of the SIGCHI Conference on Human Factors in Computing Systems*, New York, NY, USA, 2008.

[45] J. J. Lin, L. Mamykina, S. Lindtner, G. Delajoux and H. B. Strub, "Fish'N'Steps: Encouraging Physical Activity with an Interactive Computer Game," in *Proceedings of the 8th International Conference on Ubiquitous Computing*, Berlin, Heidelberg, 2006.

[46] G. Adomavicius and A. Tuzhilin, "Toward the next generation of recommender systems: a survey of the state-of-the-art and possible extensions," *Knowledge and Data Engineering, IEEE Transactions on,* vol. 17, no. 6, pp. 734-749, June 2005.

[47] J. Heckhausen and C. S. Dweck, Motivation and self-regulation across the life span, Cambridge University Press, 1998.

[48] T. C. Leonard, "Richard H. Thaler, Cass R. Sunstein, Nudge: Improving decisions about health, wealth, and happiness," *Constitutional Political Economy,* vol. 19, no. 4, pp. 356-360, 2008.